\documentclass[11pt]{article}
\usepackage{graphicx}
\begin{document}

\centerline{\bf\Large  Study on Strong Decays of $D_{sJ}(2632)$ }

\vspace{1cm}

Hong-Wei Ke$^1$, Yan-Ming Yu$^1$,  Yi-Bing Ding$^2$, Xin-Heng
Guo$^3$, Hong-Ying Jin$^4$, Xue-Qian Li$^1$, Peng-Nian Shen$^5$
and Guo-Li Wang$^6$

\vspace{0.5cm}

1. Department of Physics, Nankai University, Tianjin 300071,
China.

2. Department of Physics,  Graduate University of Chinese Academy
of Sciences, Beijing, 100049, China.

3. Institute of Low Energy Nuclear Physics, Beijing Normal
University, Beijing 100875, China.

4. Institute of Modern Physics, Zhejiang University, Hangzhou,
310027, China.

5. Institute of High Energy Physics, Chinese Academy of Sciences,
P.O. Box 918-4, Beijing 100049, China.

6. Department of Physics, Harbin Institute of Technology, Harbin,
150006, China

\vspace{1cm}

\begin{center}
\begin{minipage}{12cm}
\noindent{Abstract}

The resonance $D_{sJ}(2632)$ observed by SELEX, has attracted
great interests and meanwhile brought up serious dispute. Its
spin-parity, so far has not finally determined and if it exists,
its quark-structure might be exotic. Following the previous
literature where $D_{sJ}(2632)$ is assumed to be a radial-excited
state of $1^-$, we consider the possibilities that it might be a
$q\bar q$ ground state of $2^+$ or the first radial-excited state
of $0^+$ $D_{sJ}(2317)$ and re-calculate its strong decay widths
in terms of the Bethe-Salpeter equation. Our results indicate that
there still is a sharp discrepancy between the theoretical
evaluation and data.

\end{minipage}
\end{center}

\vspace{0.2cm}

\section{Introduction}

In 2004, the SELEX collaboration reported that a charmed meson
$D_{sJ}(2632)$ with narrow width was observed \cite{2632}, its
mass is $2632.5\pm1.7MeV$ at 90\% CL. This state was observed in
two channels ${D_s}^{+} \eta$ and $D^{0}K^{+}(D^{+}K^{0})$ which
are OZI allowed processes. The measured ratio of their  branching
ratios $R=\Gamma(D_{sJ}^{+}\rightarrow
D^{0}K^{+})/\Gamma(D_{sJ}^{+}\rightarrow {D_s}^{+}
\eta)=0.14\pm0.06$ shows a strange pattern. Because the phase
space available for $D^{0}K^{+}$ is almost 1.5 times larger than
that for ${D_s}^{+} \eta$, but the final products possess the same
quark contents, therefore one is tempted to believe that this
discrepancy implies that the quark structure of $D_{sJ}(2632)$
might be exotic. To eventually confirm the allegation, one needs
to carefully seek for solutions in the traditional theoretical
framework. In fact, the spin-parity of the resonance
$D_{sJ}(2632)$ has not finally determined yet. According to the
final states, it may be $0^+$, $1^-$ or even $2^+$ \cite{jp}which
can decay into two pseudoscalar mesons via S-, P- and D-waves
respectively.

Actually, before considering the exotic structure, such as hybrid,
four-quark state etc., one should seriously investigate if it can
be embedded in the frame of regular mesons which contain only one
quark and one anti-quark\cite{Eef,other,Simonov}. Chang et
al.\cite{Wang} assumed that $D_{sJ}(2632)$ is the first radial
excited state of $D_s^*$ and has spin-parity as $1^-$, and then
they evaluated its decay widths in terms of the Bethe-Salpeter
equation. Here we consider alternative possibilities that
$D_{sJ}(2632)$ may be $0^+$ or $2^+$ resonances of $c\bar s$.
Namely, we suppose that $D_{sJ}(2632)$ is the first radial excited
state of $D_{sJ}(2317)$ whose spin-parity is well determined by
several collaborations\cite{2317}, or the $2^+$ radial ground
state of $c\bar s$. Then following Chang et al.\cite{Wang}, we
also evaluate the decay widths of $D_{sJ}^{+}(2632)\rightarrow
D^{0}K^{+})$ and $(D_{sJ}^{+}(2632)\rightarrow {D_s}^{+} \eta)$ by
assuming it to be of a radial excited state of $0^+$ or a radial
ground state of $2^+$ with only $c\bar s$ quark contents.

In the derivation, we have also considered two approaches for the
$0^+$ case. First, we assume that it is the first radial excited
state of the observed $D_{sJ}(2317)$  whose mass is 2317 MeV, and
then we obtain a mass of the excited state as $2700\pm 20MeV$ (see
the following text for details). Alternatively, if we assume 2632
MeV as the mass of an $0^+$ radial excited state, then we obtain
the mass of its corresponding radial ground state of $0^+$ as
$2245\pm 25MeV$. That is in analog to the approach of
ref.\cite{Wang}, where the authors calculated the mass of the
first radial excited state of $D_s^*$ as $2658\pm 15$ MeV. In the
second approach, the newly obtained mass $2245\pm 25MeV$ obviously
deviates from the observed 2317 MeV which is well measured,
therefore unless a new resonance $D_{sJ}$ of $0^+$ were
experimentally observed, this scenario is not favored by the
present data. However, considering the experimental errors, it may
still be possible, and we will further discuss it in the last
section.

This work is organized as follows, after this introduction, we
discuss all the aforementioned  possibilities and by solving the
B-S equation, we obtain the mass spectrum and the OZI-allowed
decay widths. In Sec.II, we deal with the case where
$D_{sJ}(2632)$ is assumed to be the radial ground state of $2^+$,
while in Sec.III, we assume that it is a radial excited state of
$0^+$. All the numerical results along with all the input
parameters are presented in the sections. The last section is
devoted to the discussions and a brief conclusion.

\section{$D_{sJ}(2632)$ as the ground state of  $2^+$ $c\bar s$}

The B-S equation with instantaneous approximation about the
$0^{-},1^{-}$ mesons has been thoroughly studied \cite{bs}.
Following the method and technical details introduced in
ref.\cite{bs,bs2}, we solve the B-S equation for the mesons of
$2^+$ under the instantaneous approximation.

Generally, the B-S wavefunction for a $2^+$ meson can be wrtten as
\cite{bs2,jin}:
\begin{eqnarray}
\varphi_{_{P_{i}}}(\textbf{q})&=&\epsilon_{ij}q^j_{\perp}
\{q^i_{\perp}
[\varphi_{1}(\textbf{q})+\gamma_{0}\varphi_{2}(\textbf{q})
+q\!\!\!\slash_{\perp}\varphi_{3}(\textbf{q})+\gamma_{0}q\!\!\!\slash_{\perp}\varphi_{4}(\textbf{q})]
+\nonumber\\&&\gamma^i[\varphi_{5}(\textbf{q})+\gamma_{0}\varphi_{6}(\textbf{q})
+q\!\!\!\slash_{\perp}\varphi_{7}(\textbf{q})]+i\epsilon^{0ilk}q_{\perp
l}\gamma_{k}\gamma_{5}\varphi_{8}(\textbf{q})\}.
\end{eqnarray}
where $\varphi_i(\textbf{q})$ is the component function,
$q_\perp=(0, \textbf{q})$, and $\textbf{q}$ is the relative
three-momentum of the quark-anti-quark in the meson,
$\epsilon^{0ilk}$ is the fully antisymmetric tensor and
$\epsilon_{ij}$ is the polarization tensor of $2^+$. For the
convenience, we redefine $\psi_{1}=\varphi_{1}$,
$\psi_{2}=\varphi_{2}$, $\psi_{3}=\textbf{q}^2\varphi_{3}$,
$\psi_{4}=\textbf{q}^2\varphi_{4}$, $\psi_{5}=\varphi_{5}$,
$\psi_{6}=-\varphi_{6}$, $\psi_{7}=\varphi_{7}$,
$\psi_{8}=\varphi_{8}$.

By the well-known constraint conditions for the projected
wavefunctions $\varphi^{+-}_{_{P_{i}}}=0$ and
${\varphi^{-+}_{_{P_{i}}}}=0$\cite{bs,bs2,Greiner}, one has
\begin{eqnarray*}
\psi_{1}(\textbf{q})=\frac{-((\omega_{1}+\omega_{2})\psi_{3}(\textbf{q})-2\omega_{2}\psi_{5}(\textbf{q}))}
{\omega_{2}m_1+\omega_{1}m_2},
\end{eqnarray*}
\begin{eqnarray*}
\psi_{2}(\textbf{q})=\frac{-(\omega_{1}-\omega_{2})(\psi_{4}(\textbf{q})-\psi_{6}(\textbf{q}))}
{\omega_{2}m_1+\omega_{1}m_2},
\end{eqnarray*}
\begin{eqnarray*}
\psi_{7}(\textbf{q})=\frac{(\omega_{1}-\omega_{2})\psi_{5}(\textbf{q})}{\omega_{2}m_1+\omega_{1}m_2},
\end{eqnarray*}
\begin{eqnarray}
\psi_{8}(\textbf{q})=-\frac{(\omega_{1}+\omega_{2})\psi_{6}(\textbf{q})}{\omega_{2}m_1+\omega_{1}m_2},
\end{eqnarray}
where $\omega_{1}=\sqrt{m^2_{1}+\mathbf{q}^2}$,
$\omega_{2}=\sqrt{m^2_{2}+\mathbf{q}^2}$, and in this text $m_1$
and $m_2$ stand as $m_c$ and $m_q$, which are masses of charm
quark and light flavor $q$ (q=u,d,s).

Thus the wavefunction of a $2^+$ meson can be further written as
\begin{eqnarray}
\varphi_{_{P_{i}}}(\textbf{q})&=&\epsilon_{ij}q^j_{\perp}
\{q^i_{\perp}
[\psi_{3}(\textbf{q})(\frac{q\!\!\!\slash_{\perp}}{\textbf{q}^2}-\frac{(\omega_1+\omega_2)}{m_2\omega_1+m_1\omega_2})
+\psi_{4}(\textbf{q})\gamma_0(\frac{q\!\!\!\slash_{\perp}}
{\textbf{q}^2}-\frac{(\omega_1-\omega_2)}{m_2\omega_1+m_1\omega_2})\nonumber\\&&
+\psi_{5}(\textbf{q})\frac{2\omega_2}{m_2\omega_1+m_1\omega_2}+
\psi_{6}(\textbf{q})\gamma_0\frac{(\omega_1-\omega_2)}{m_2\omega_1+m_1\omega_2}]
\nonumber\\&&+\gamma^i[\psi_{5}(\textbf{q})(1+\frac{q\!\!\!\slash_{\perp}(\omega_1-\omega_2)}{m_2\omega_1+m_1\omega_2})
-\gamma_{0}\psi_{6}(\textbf{q})]\nonumber\\&&-i\epsilon^{0ilk}q_{\perp
l}\gamma_{k}\gamma_{5}\psi_{6}(\textbf{q})
\frac{(\omega_1+\omega_2)}{m_2\omega_1+m_1\omega_2}\}.
\end{eqnarray}

Then, we obtain an equation group which contains four mutually
coupled equations, the detailed expressions are collected in
appendix.

\begin{figure}[t]
\begin{center}
\begin{tabular}{ccc}
\scalebox{0.7}{\includegraphics{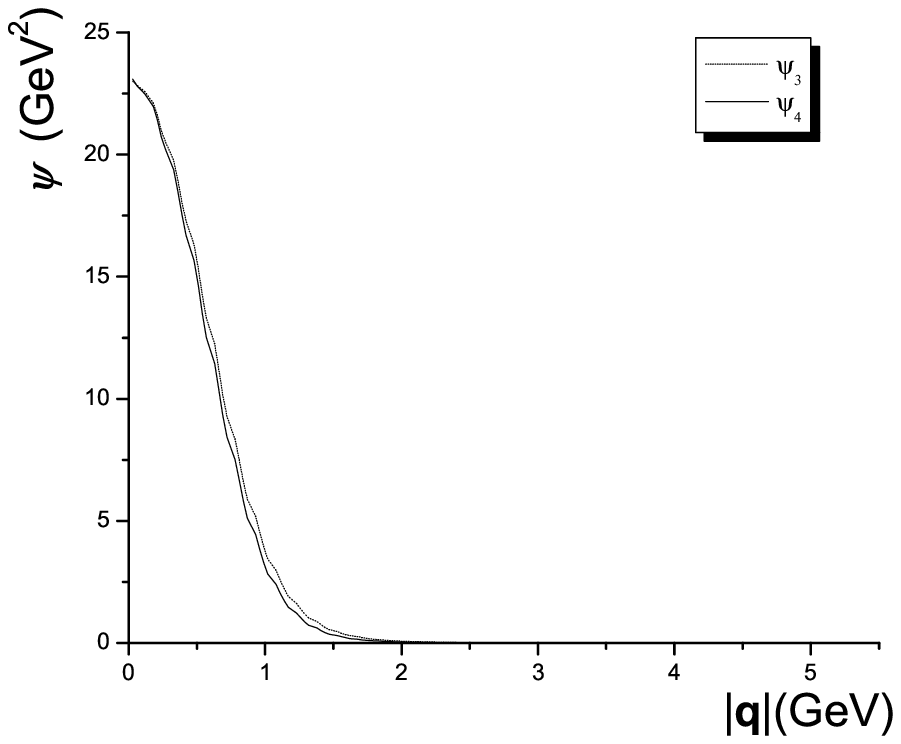}}&&\scalebox{0.7}{\includegraphics{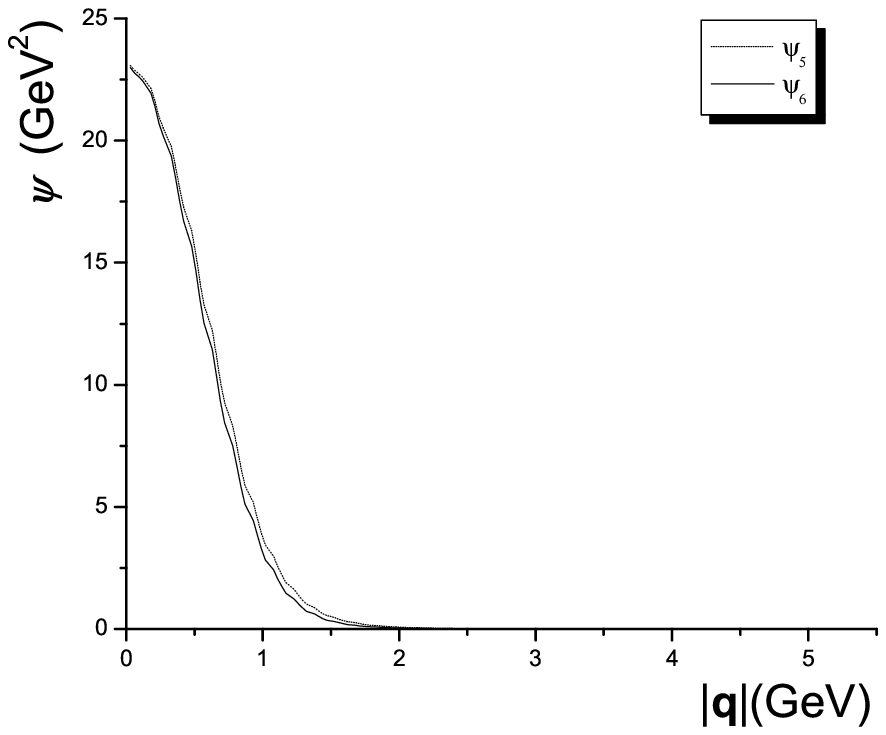}}
\\

\end{tabular}
\end{center}
\caption{The component functions of $D_{sJ}(2632)$ which is
assumed to be the radial ground state of $2^+$ $c\bar q$,
$\psi_{3}$ , $\psi_{4}$, $\psi_{5}$ , $\psi_{6}$.}\label{wave1}
\end{figure}

In this work we adopt the values given in \cite{Wang} for the
concerned parameters, but only change $V_0$ to obtain the mass of
$2632\pm 16MeV$ for the ground state of $2^+$. By solving the
equation group, numerical solutions for the component functions
$\psi_{3}$, $\psi_{4}$, $\psi_{5}$, $\psi_{6}$ are achieved,
 these functions are shown in Fig.(\ref{wave1}). Actually,
$\psi_3\approx\psi_4$, and $\psi_5\approx \psi_6$, therefore, in
the figure they seem to overlap together.

Now, we can use the formula given by ref.\cite{Wang} to evaluate
the widths of the strong decays.

\begin{eqnarray}\label{kd}
\Gamma=\frac{|\mathbf{P}_{f1}|}{8\pi M^2}|T|^2,
\end{eqnarray}
where $M$ is the mass of the initial meson $D_{sJ}(2632)$,
$\mathbf{P}_{f1}$ is the three momentum of the produced mesons
$D$( or $D_s$) in the center of mass frame of $D_{sJ}(2632)$. For
$D_{sJ}^{+}\rightarrow D^{0}K^{+}$ and $D_{sJ}^{+}\rightarrow
D^{+}K^{0}$, the matrix element $T$ is
\begin{eqnarray}\label{t1}
T=\frac{P^\mu_{f2}}{f_{K}}\int\frac{d\textbf{q}}{(2\pi)^3}Tr[\overline{
\varphi}^{++}_{_{P_{f1}}}(\textbf{q}-\frac{m_1}{m_1+m_2}\textbf{p}_{f1})
\frac{P\!\!\!\!\slash_i}{M}{\varphi}^{++}_{_{P_{i}}}(\textbf{q})
\gamma_\mu\gamma_5].
\end{eqnarray}

For $D_{sJ}^{+}\rightarrow {D_s}^{+} \eta$, it is
\begin{eqnarray}\label{t2}
T=P^\mu_{f2}[\frac{-2M^2_{\eta}cos\theta}{\sqrt{6}M^2_{\eta_8}f_{\eta_8}}+\frac{M^2_{\eta}sin\theta}{\sqrt{3}M^2_{\eta_0}
f_{\eta_0}}]\int\frac{d\textbf{q}}{(2\pi)^3}Tr[\overline{
\varphi}^{++}_{_{P_{f1}}}(\textbf{q}-\frac{m_1}{m_1+m_2}
\textbf{p}_{f1})\frac{P\!\!\!\!\slash_i}{M}{\varphi}^{++}_{_{P_{i}}}(\textbf{q})
\gamma_\mu\gamma_5].
\end{eqnarray}

Here $\mathbf{q}$ is the inner relative three-momentum in the
initial meson $D_{sJ}(2632)$, $P_{f2}$ is the four-momentum of the
produced meson $K$ (or $\eta$), $P_{i}$ is the four-momentum of
$D_{sJ}(2632)$, $M_\eta$ is the mass of $\eta$.
 ${\varphi}^{++}_{_{P_{i}}}$ and ${\varphi}^{++}_{_{P_{f_1}}}$ is the
positive-energy wavefunction of the initial or final meson, and
$\overline{\varphi}^{++}_{_{P_{f1}}}=-\gamma_{0}(\varphi^{++}_{_{P_{f1}}})^{+}\gamma_{0}$.
The factor in eq.(\ref{t2})
$[\frac{-2M^2_{\eta}cos\theta}{\sqrt{6}M^2_{\eta_8}f_{\eta_8}}+\frac{M^2_{\eta}sin\theta}{\sqrt{3}M^2_{\eta_0}
f_{\eta_0}}]$ takes into account the $\eta-\eta'$ mixing, the
readers are recommended to refer to ref.\cite{Wang} for some
details. $f_{K}$ is the decay constant of $K$ meson, $f_{\eta_8}$,
$f_{\eta_0}$ are the decay constants of $\eta_8$ and $\eta_0$
respectively.

As the decay products are pseudoscalar mesons, their
positive-energy wavefunctions are\cite{bs}
\begin{eqnarray}
\varphi^{++}_{_{P_{f1}}}(\textbf{q})=\frac{M_{f1}}{2}(\varphi_{1}(\textbf{q})+\varphi_{2}(\textbf{q})\frac{m_1m_2}
{\omega_{1}\omega_{2}})[\frac{\omega_{1}+\omega_{2}}{m_1+m_2}+\\\gamma_{0}-\frac{q\!\!\!\slash_\perp(m_1-m_2)}
{m_2\omega_{1}+m_1\omega_{2}}+\frac{\gamma_{0}q\!\!\!\slash_\perp(m_1-m_2)}{m_2\omega_{1}+m_1\omega_{2}}]
\gamma_{5}.
\end{eqnarray}

The relation of the positive-energy wavefunction of the initial
meson and the its wavefunction $\varphi_{P_{i}}(\textbf{q})$ reads
\cite{Wang}
\begin{eqnarray}\label{positive}
\varphi^{++}_{_{P_{i}}}(\textbf{q})=\frac{1}{2\omega_{1}}(\omega_{1}\gamma_{0}+m_{1}+q\!\!\!\slash_\perp)
\gamma_{0}\varphi_{_{P_{i}}}(\textbf{q})\gamma_{0}\frac{1}{2\omega_{2}}(\omega_{2}\gamma_{0}-m_2-q\!\!\!\slash_\perp).
\end{eqnarray}

Using the wavefunctions obtained by solving the equations, we
evaluate the partial widths as
\begin{eqnarray}
\Gamma(D_{sJ}^{+}\rightarrow D^{0}K^{+})&=&2.10\pm 0.30MeV,\\
\Gamma(D_{sJ}^{+}\rightarrow D^{+}K^{0})&=&2.22\pm 0.31MeV,\\
\Gamma(D_{sJ}^{+}\rightarrow {D_s}^{+} \eta)&=&0.23\pm 0.02MeV.
\end{eqnarray}

The corresponding ratio of the branching ratios is
\begin{eqnarray}
\Gamma(D_{sJ}^{+}\rightarrow
D^{0}K^{+})/\Gamma(D_{sJ}^{+}\rightarrow {D_s}^{+} \eta)
\approx\Gamma(D_{sJ}^{+}\rightarrow
D^{+}K^{0})/\Gamma(D_{sJ}^{+}\rightarrow {D_s}^{+} \eta)
\approx9.2\pm0.9.
\end{eqnarray}

If we assume that the observed $D_{sJ}(2632)$ is a radial ground
state of $2^+$, the obtained total width is consistent with the
data, but the ratio of the branching ratios obviously differs from
the observation.

In the numerical computations, we choose the input parameters as
ref.\cite{Wang}, $m_c=1755.3MeV$, $m_s=487MeV$, $m_d=311MeV$,
$m_u=305MeV$, $f_\pi=130.7MeV$, $f_{\eta_8}=1.26f_\pi$,
$f_{\eta_8}=1.07f_\pi$, $f_{K}=159.8MeV$, $M_{\eta_8}=564.3MeV$,
$M_{\eta_0}=948.1MeV$ and $\theta=-9.95^\circ$. The theoretical
uncertainties are estimated by varying all the parameters
simultaneously within $\pm5\%$.

\section{$D_{sJ}(2632)$ as the first radial excited state of $0^+$ $c\bar s$}

In our earlier work, we obtained the expression of the
wavefunction for $0^+$ diquark \cite{yu}, for a $0^+$ meson which
is composed of a quark and an antiquark, the B-S equation is the
same, but the integration kernel is different.

The wavefunction of a $0^+$ scalar meson is written as
\begin{eqnarray}
\varphi_{_{_{P_{i}}}}(\textbf{q})=
[\varphi_{1}(\textbf{q})+\gamma_{0}\varphi_{2}(\textbf{q})
+q\!\!\!\slash_{\perp}\varphi_{3}(\textbf{q})+\gamma_{0}q\!\!\!\slash_{\perp}\varphi_{4}(\textbf{q})].
\end{eqnarray}

For the convenience, we redefine $\psi_{1}=\varphi_{1}$,
$\psi_{2}=\varphi_{2}$,
$\psi_{3}=|\textbf{q}|\varphi_{3}$,$\psi_{4}=|\textbf{q}|\varphi_{4}$

By the constraint condition of projected wave function, one has
\begin{eqnarray*}
\psi_{1}(\textbf{q})=-\frac{(\omega_{1}-\omega_{2})\psi_{4}(\textbf{q})|\textbf{q}|}{\omega_{2}m_1+\omega_{1}m_2},
\end{eqnarray*}
\begin{eqnarray}
\psi_{3}(\textbf{q})=\frac{(\textbf{q}^2-m_1m_2-\omega_{1}\omega_{2})\psi_{2}(\textbf{q})}{(m_1+m_2)|\textbf{q}|}.
\end{eqnarray}

With the constraints, one can further write the wavefunction as
\begin{eqnarray}
\varphi_{_{P_{i}}}(\textbf{q})=
\psi_{2}(\textbf{q})[\gamma_{0}+q\!\!\!\slash_{\perp}\frac{\textbf{q}^2-m_1m_2-\omega_1\omega_2}
{(m_1+m_2)\textbf{q}^2}]+\psi_{4}(\textbf{q})(\frac{\gamma_{0}q\!\!\!\slash_{\perp}}{|\textbf{q}|}-
\frac{(\omega_1-\omega_2)|\textbf{q}|}{m_1\omega_2+m_2\omega_1}).
\end{eqnarray}

Substituting this wavefunction into (\ref{positive}), we obtain
the positive-energy wavefunction of $0^+$ scalar meson.

\begin{figure}[t]
\begin{center}
\begin{tabular}{ccc}
\scalebox{0.7}{\includegraphics{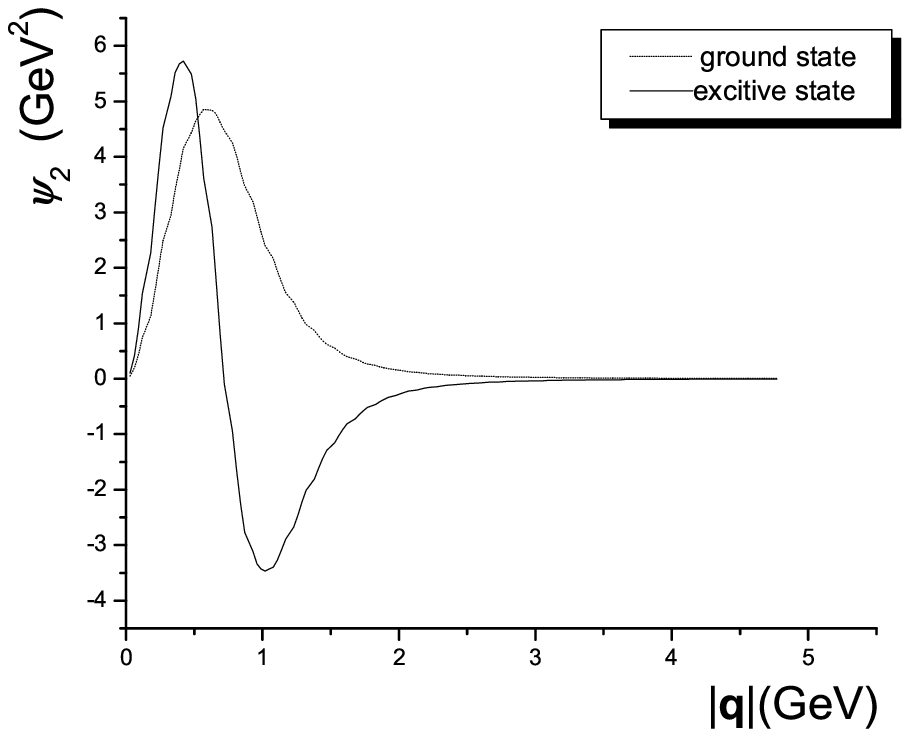}}&&\scalebox{0.7}{\includegraphics{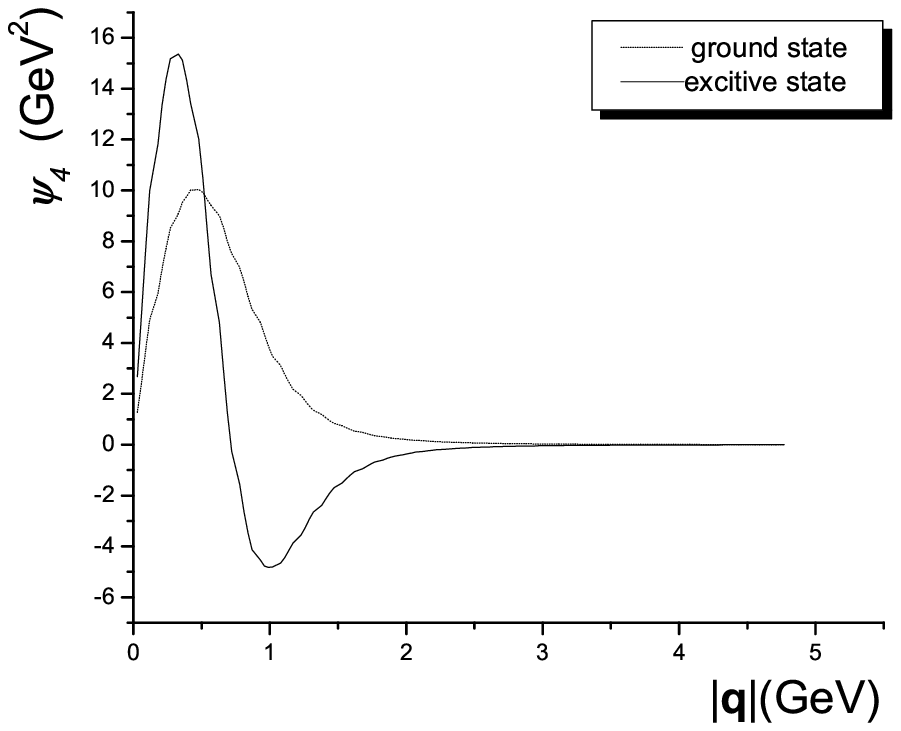}}
\\
\\
\end{tabular}
\end{center}
\caption{Case 1. we adjust parameters when 2317 MeV serves as the
basic input parameter,the curves correspond to the component
wavefunctionss of the $0^+$ ground state and first radial excited
state $\psi_{2}$, $\psi_{4}$ respectively. }\label{wave2}
\end{figure}

We can simplify the B-S equation  and obtain an equation group
which only contains two coupled equations, the detailed
expressions are presented in the appendix.

As we indicated in the introduction, we take two approaches.

1. Adjusting parameters while 2317 MeV is taken as the basic input
parameter, namely the observed meson (may correspond to
$D_{sJ}(2632)$) is supposed to be the first excited state of the
well measured $D_{sJ}(2317)$:

Adopting the parameter given in the ref.\cite{Wang}, but varying
$V_0$ to fit the mass of the ground state of $0^+$ which is set as
2317 MeV, we obtain the mass of the first radial excited state as
$2700\pm 20MeV$, and the corresponding component wavefunctions.
$\psi_{2}$, $\psi_{4}$ are shown in Fig.\ref{wave2}.

\begin{figure}[t]
\begin{center}
\begin{tabular}{ccc}
\scalebox{0.7}{\includegraphics{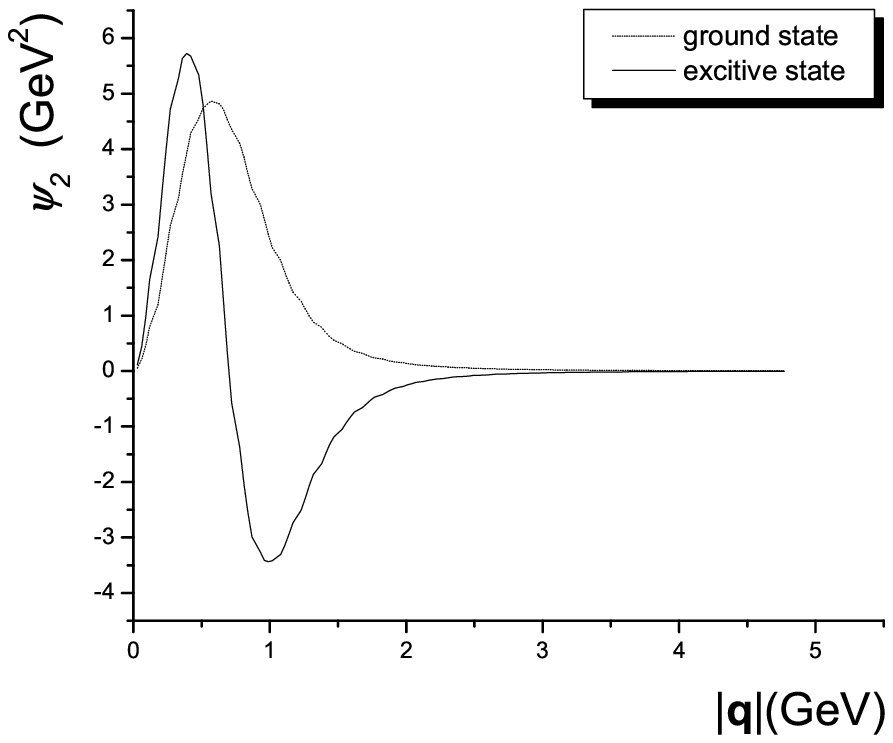}}&&\scalebox{0.7}{\includegraphics{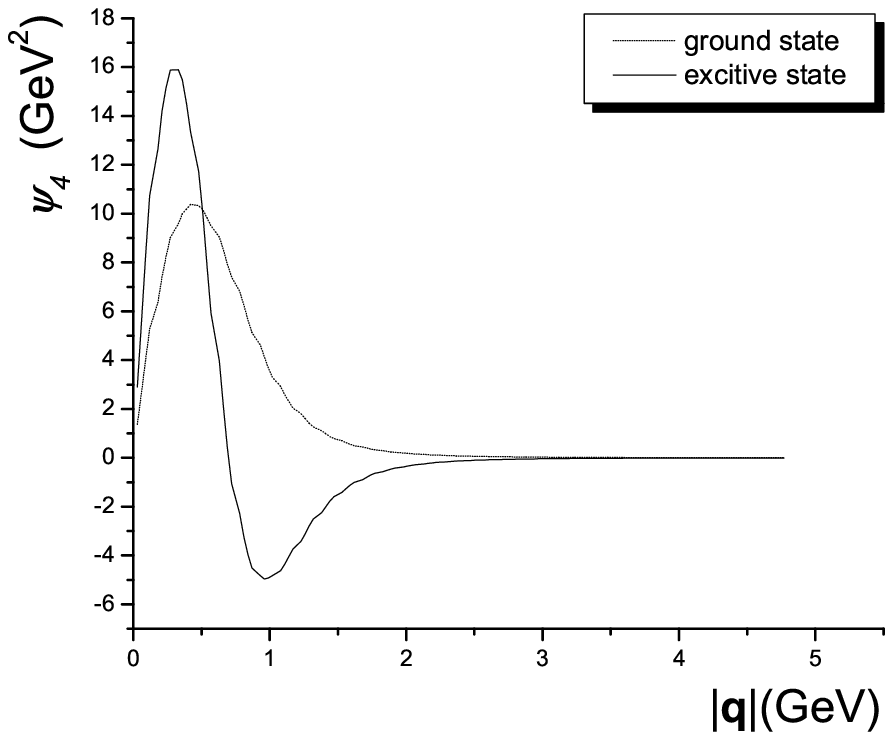}}
\\
\end{tabular}
\end{center}
\caption{Instead, the corresponding component wavefunctions of the
$0^+$ ground state and first radial excited state $\psi_{2}$,
$\psi_{4}$, as we take 2632 MeV as the input parameter of the mass
of the first radial excited state to adjust other parameters }
\label{wave3}
\end{figure}
Substituting the wavefunctions into eq.(\ref{kd}), we obtain the
numerical values

\begin{eqnarray}
\Gamma(D_{sJ}^{+}\rightarrow D^{0}K^{+})&=&9.58\pm 1.33MeV,\\
\Gamma(D_{sJ}^{+}\rightarrow D^{+}K^{0})&=&9.36\pm 1.30MeV,\\
\Gamma(D_{sJ}^{+}\rightarrow {D_s}^{+} \eta)&=&3.64\pm 0.31MeV.
\end{eqnarray}

The corresponding ratio of the branching ratios would be
\begin{eqnarray}
\Gamma(D_{sJ}^{+}\rightarrow
D^{0}K^{+})/\Gamma(D_{sJ}^{+}\rightarrow {D_s}^{+} \eta)
\approx\Gamma(D_{sJ}^{+}\rightarrow
D^{+}K^{0})/\Gamma(D_{sJ}^{+}\rightarrow {D_s}^{+} \eta)
\approx2.6\pm0.3.
\end{eqnarray}

2. Numerical results while 2632 MeV is taken as the input
parameter for the mass of the first radial excited state of an
undiscovered  $0^+$ meson (it might be the measured $D_{sJ}(2317)$
if the experimental errors are indeed large).

While we adjust $V_0$ to fit 2632 MeV which is taken as the input
parameter for the mass of the first radial excited state, we
obtain the mass of the ground state of $0^+$ as $2245\pm 25MeV$.
The corresponding component wavefunctions $\phi_{2}$, $\phi_{4}$
are shown in Fig. \ref{wave3}.

Substituting the wavefunctions into eq.(\ref{kd}), we obtain
\begin{eqnarray}
\Gamma(D_{sJ}^{+}\rightarrow D^{0}K^{+})&=&7.58\pm 1.06MeV,\\
\Gamma(D_{sJ}^{+}\rightarrow D^{+}K^{0})&=&7.25\pm 1.01MeV,\\
\Gamma(D_{sJ}^{+}\rightarrow {D_s}^{+} \eta)&=&2.37\pm 0.20MeV.
\end{eqnarray}
and the corresponding ratio of the branching ratios
\begin{eqnarray}
\Gamma(D_{sJ}^{+}\rightarrow
D^{0}K^{+})/\Gamma(D_{sJ}^{+}\rightarrow {D_s}^{+} \eta)
\approx\Gamma(D_{sJ}^{+}\rightarrow
D^{+}K^{0})/\Gamma(D_{sJ}^{+}\rightarrow {D_s}^{+} \eta)
\approx3.1\pm0.4.
\end{eqnarray}

\section{Discussions and conclusion}

In terms of the B-S equation, we evaluate the mass spectra and the
decay rates when assuming $D_{sJ}(2632)$ observed by the SELEX
collaboration to be a $0^+$ or $2^+$  meson of $c\bar s$.

Our observation is that if $D_{sJ}(2632)$ is a $2^+$ $c\bar s$
state, the total width obtained in terms of the B-S equation can
be consistent with data, but the predicted branching ratios
obviously conflict with data. If it is a $0^+$ $c\bar s$ state,
our results can be summarized as follows. In approach (1), we
adjust $V_0$ to fit the mass 2317 MeV which is the mass of the
$0^+$ ground state of $c\bar s$, and obtain the mass of the first
radial excited state as $2700\pm 20MeV$. Instead, in approach (2),
we adjust $V_0$ to fit 2632 MeV which is taken as the mass of the
first radial excited state of $0^+$ $c\bar s$, then we get the
mass of the ground state as $2245\pm 25MeV$. It is noted that the
wavefunctions obtained in the two cases are very close as shown in
Figs.\ref{wave2} and \ref{wave3}. The decay rates calculated in
approach (1) are a bit larger than that in approach (2) due to a
larger phase space. Thus in approach (2), the obtained total width
is consistent with data, but not the ratio of branching ratios
within the experimental tolerance range, whereas in approach (1),
both the total width and the ratio do not coincide with the
present data.

Moreover, if $D_{sJ}(2632)$ is  a $1^-$ $c\bar s$ vector meson and
is the first radial excited state of $D_s^*$, as Chang et al.
evaluated \cite{Wang}, the ratio of branching ratios is also
inconsistent with data.

Definitely,  the calculations are model-dependent, especially the
linear confinement part of the kernel in the B-S equation is
phenomenologically introduced and the coefficient $\kappa$ is
determined by fitting data. Thus the numerical results obtained in
this theoretical framework cannot be very accurate, a factor of as
large as $2\sim 3$ may be expected, however, the order of
magnitude and the qualitative behavior of the solution do not
change.

Therefore, there may be some possible explanations for obvious
discrepancy between data and theoretical result. First, from the
theoretical aspect, the discrepancy may indicate that
$D_{sJ}(2632)$ possess a large exotic component, namely it may be
a four-quark state\cite{tetraquark}, molecular state, or a
hybrid\cite{hybridandmolecular,Dai,Chao}.

Another possibility is, as Swanson et al. suggested
\cite{Swanson,close}, it could be the "artefact" of experiments.
Moreover, the other prestigious experimental groups
\cite{Swanson,notfound} do not see evidence of $D_{sJ}(2632)$,
therefore, its existence is still suspicious. Further and more
precise measurements are necessary to clarify this mystery, great
efforts from both sides of  theory and experiments must be made.

\vspace{1cm}

\noindent Acknowledgments:

We thank C.-H. Chang for helpful discussions. This work is partly
supported by the National Natural Science Foundation of China
(NNSFC).

\appendix{\textbf{Appendix}}
\section{ $2^+$ }
\textbf{normalization condition}
\begin{eqnarray}
\int\frac{d\mathbf{q}}{2\pi^3}\frac{16\textbf{q}^2(2\psi_3\psi_4+2\psi_3\psi_6-2\psi_4\psi_5-5\psi_5\psi_6)
\omega_1\omega_2}{3(\omega_1 m_2+\omega_2 m_1)}=2P_{i0}
\end{eqnarray}
 \textbf{Coupled equations }
\begin{eqnarray}
&&\frac{(M-\omega_1-\omega_2)4\textbf{q}^4((m_1+m_2)\psi_{4}(\mathbf{q})+(\omega_1+\omega_2)\psi_{3}(\mathbf{q})
-(m_1+m_2)\psi_{6}(\mathbf{q})-(\omega_1+\omega_2)\psi_{5}(\mathbf{q}))}{3(m_2\omega_1+m_1\omega_2)} \nonumber\\
&&=\int\frac{d\mathbf{k}}{(2\pi)^3}\big \{ -3(V_s-V_v)\big[(m_1+m_2)\psi_{3}(\mathbf{q})
+(\omega_1+\omega_2)\psi_{4}(\mathbf{q})   \big] (m_2\omega_{1k}+m_1\omega_{2k})(\mathbf{k}\cdot \mathbf{q})^3/\textbf{k}^2\nonumber\\
&&-3(V_s+V_v)\big [
-\psi_{6}(\mathbf{q})(m_2\omega_1\omega_{1k}-m_1\omega_2\omega_{1k}
-m_2\omega_1\omega_{2k}
-m_1\omega_2\omega_{2k}) +\nonumber\\
&&\psi_{5}(\mathbf{q})(\omega_{1k}\mathbf{q}^2+\omega_{2k}\mathbf{q}^2-m_1m_2\omega_{1k}+\omega_{1}\omega_{2}\omega_{1k}
-m_1m_2\omega_{2k}+\omega_{1}\omega_{2}\omega_{2k})+\nonumber\\
&&\psi_{4}(\mathbf{q})(m_1\omega_2-m_2\omega_1)(\omega_{2k}-\omega_{1k})+\nonumber
\psi_{3}(\mathbf{q})(-\mathbf{q}^2+m_1m_2-\omega_1\omega_2)(\omega_{1k}+\omega_{2k})\big](\mathbf{k}\cdot
\mathbf{q})^2
  \nonumber\\&&\mathbf{q}^2(V_s-V_v)\big[\psi_{4}(\mathbf{q})(\omega_1+\omega_2)+\psi_{3}(\mathbf{q})(m_1+m_2)
  +2\psi_{5}(\mathbf{q})(m_1+m_2) \nonumber\\
&&+2\psi_{6}(\mathbf{q})(\omega_1+\omega_2)\big](m_2\omega_{1k}+m_1\omega_{2k})\mathbf{k}\cdot \mathbf{q}+\nonumber\\
&&\mathbf{k}^2\mathbf{q}^2(V_s+V_v)\big [-
\psi_{6}(\mathbf{q})(m_2\omega_1\omega_{1k}-m_1\omega_2\omega_{1k}
-m_2\omega_1\omega_{2k}
-m_1\omega_2\omega_{2k}) +\nonumber\\
&&\psi_{5}(\mathbf{q})(\omega_{1k}\mathbf{q}^2+\omega_{2k}\mathbf{q}^2-m_1m_2\omega_{1k}+\omega_{1}\omega_{2}\omega_{1k}
-m_1m_2\omega_{2k}+\omega_{1}\omega_{2}\omega_{2k})+\nonumber\\
&&\psi_{4}(\mathbf{q})(m_1\omega_2-m_2\omega_1)(\omega_{2k}-\omega_{1k})+\nonumber\\
&&\psi_{3}(\mathbf{q})(-\mathbf{q}^2+m_1m_2-\omega_1\omega_2)(\omega_{1k}+\omega_{2k})\big]
\big\} /[3\omega_1\omega_2(m_2\omega_{1k}+m_1\omega_{2k})],
\end{eqnarray}

\begin{eqnarray}
&&\frac{(M+\omega_1+\omega_2)4\textbf{q}^4((m_1+m_2)\psi_{4}(\mathbf{q})-(\omega_1+\omega_2)\psi_{3}(\mathbf{q})
-(m_1+m_2)\psi_{6}(\mathbf{q})+(\omega_1+\omega_2)\psi_{5}(\mathbf{q}))}{3(m_2\omega_1+m_1\omega_2)} \nonumber\\
&&=\int\frac{d\mathbf{k}}{(2\pi)^3}\big \{ -3(V_s-V_v)\big[(m_1+m_2)\psi_{3}(\mathbf{q})
-(\omega_1+\omega_2)\psi_{4}(\mathbf{q})   \big] (m_2\omega_{1k}+m_1\omega_{2k})(\mathbf{k}\cdot \mathbf{q})^3/\textbf{k}^2\nonumber\\
&&-3(V_s+V_v)\big [
\psi_{6}(\mathbf{q})(m_2\omega_1\omega_{1k}-m_1\omega_2\omega_{1k}
-m_2\omega_1\omega_{2k}
-m_1\omega_2\omega_{2k}) +\nonumber\\
&&\psi_{5}(\mathbf{q})(\omega_{1k}\mathbf{q}^2+\omega_{2k}\mathbf{q}^2-m_1m_2\omega_{1k}+\omega_{1}\omega_{2}\omega_{1k}
-m_1m_2\omega_{2k}+\omega_{1}\omega_{2}\omega_{2k})-\nonumber\\
&&\psi_{4}(\mathbf{q})(m_1\omega_2-m_2\omega_1)(\omega_{2k}-\omega_{1k})+\nonumber
\psi_{3}(\mathbf{q})(-\mathbf{q}^2+m_1m_2-\omega_1\omega_2)(\omega_{1k}+\omega_{2k})\big](\mathbf{k}\cdot
\mathbf{q})^2
  \nonumber\\&&\mathbf{q}^2(V_s-V_v)\big[-\psi_{4}(\mathbf{q})(\omega_1+\omega_2)+\psi_{3}(\mathbf{q})(m_1+m_2)
  +2\psi_{5}(\mathbf{q})(m_1+m_2) \nonumber\\
&&-2\psi_{6}(\mathbf{q})(\omega_1+\omega_2)\big](m_2\omega_{1k}+m_1\omega_{2k})\mathbf{k}\cdot \mathbf{q}+\nonumber\\
&&\mathbf{k}^2\mathbf{q}^2(V_s+V_v)\big [
\psi_{6}(\mathbf{q})(m_2\omega_1\omega_{1k}-m_1\omega_2\omega_{1k}
-m_2\omega_1\omega_{2k}
-m_1\omega_2\omega_{2k}) +\nonumber\\
&&\psi_{5}(\mathbf{q})(\omega_{1k}\mathbf{q}^2+\omega_{2k}\mathbf{q}^2-m_1m_2\omega_{1k}+\omega_{1}\omega_{2}\omega_{1k}
-m_1m_2\omega_{2k}+\omega_{1}\omega_{2}\omega_{2k})-\nonumber\\
&&\psi_{4}(\mathbf{q})(m_1\omega_2-m_2\omega_1)(\omega_{2k}-\omega_{1k})+\nonumber\\
&&\psi_{3}(\mathbf{q})(-\mathbf{q}^2+m_1m_2-\omega_1\omega_2)(\omega_{1k}+\omega_{2k})\big]
\big\} /[3\omega_1\omega_2(m_2\omega_{1k}+m_1\omega_{2k})],
\end{eqnarray}

\begin{eqnarray}
&&\frac{(M-\omega_1-\omega_2)}{3(m_2\omega_1+m_1\omega_2)}2\mathbf{q}^2\big[-5\psi_{5}(\mathbf{q})(m_1\omega_2+m_2\omega_1)
+  2\psi_{3}(\mathbf{q})(m_1\omega_2+m_2\omega_1)-\nonumber\\
&&\psi_{6}(\mathbf{q})(\mathbf{q}^2+5m_1\omega_2+5m_2\omega_1)+2\psi_{4}(\mathbf{q})
(-q^2+m_1m_2+\omega_1\omega_2) \big] \nonumber\\
&&=\int\frac{d\mathbf{k}}{(2\pi)^3}\big \{ 12(V_s-V_v)\psi_{3}(\mathbf{q})
(m_2\omega_{1k}+m_1\omega_{2k})(\mathbf{k}\cdot \mathbf{q})^3/\textbf{k}^2+3(V_s+V_v)\nonumber\\
&&\big [-
\psi_{6}(\mathbf{q})(5\omega_1\omega_{1k}+\omega_2\omega_{1k}+\omega_1\omega_{2k}
+\omega_2\omega_{2k}) +\psi_{5}(\mathbf{q})(-5m_1\omega_{1k}+m_2\omega_{1k}+m_1\omega_{2k}-5m_{2}\omega_{2k})\nonumber\\
&&+2\psi_{4}(\mathbf{q})(\omega_1-\omega_2)(\omega_{1k}-\omega_{2k})+\psi_{3}(\mathbf{q})
(-\mathbf{q}^2+m_1m_2-\omega_1\omega_2)(\omega_{1k}+\omega_{2k})\big](\mathbf{k}\cdot
\mathbf{q})^2
  \nonumber\\&&2(V_s-V_v)\big[-\psi_{6}(\mathbf{q})(5m_2\omega_1+5m_1\omega_2)+2\psi_{4}(\mathbf{q})(m_2\omega_1+m_1\omega_2)
   +\psi_{5}(\mathbf{q})(-\mathbf{q}^2-5m_1m_2-5\omega_1\omega_2) \nonumber\\
&&+2\psi_{3}(\mathbf{q})(2\mathbf{q}^2+m_1m_2+\omega_1\omega_2)\big](m_2\omega_{1k}+m_1\omega_{2k})\mathbf{k}\cdot \mathbf{q}-\nonumber\\
&&\mathbf{k}^2\mathbf{q}^2(V_s+V_v)\big [
-\psi_{6}(\mathbf{q})(5\omega_1\omega_{1k}+\omega_2\omega_{1k}+\omega_1\omega_{2k}
+\omega_2\omega_{2k}) +\nonumber\\
&&\psi_{5}(\mathbf{q})(-5m_1\omega_{1k}+m_2\omega_{1k}+m_1\omega_{2k}-5m_{2}\omega_{2k})
+2\psi_{4}(\mathbf{q})(\omega_1-\omega_2)(\omega_{1k}-\omega_{2k})+\nonumber\\
&&\psi_{3}(\mathbf{q})(-\mathbf{q}^2+m_1m_2-\omega_1\omega_2)(\omega_{1k}+\omega_{2k})\big]\big\}
/[6\omega_1\omega_2(m_2\omega_{1k}+m_1\omega_{2k})],
\end{eqnarray}

\begin{eqnarray}
&&\frac{(M+\omega_1+\omega_2)}{3(m_2\omega_1+m_1\omega_2)}2\mathbf{q}^2\big[-5\psi_{5}(\mathbf{q})(m_1\omega_2+m_2\omega_1)
+  2\psi_{3}(\mathbf{q})(m_1\omega_2+m_2\omega_1)+\nonumber\\
&&\psi_{6}(\mathbf{q})(\mathbf{q}^2+5m_1\omega_2+5m_2\omega_1)-2\psi_{4}(\mathbf{q})
(-q^2+m_1m_2+\omega_1\omega_2) \big] \nonumber\\
&&=\int\frac{d\mathbf{k}}{(2\pi)^3}\big \{-12(V_s-V_v)\psi_{3}(\mathbf{q})
(m_2\omega_{1k}+m_1\omega_{2k})(\mathbf{k}\cdot \mathbf{q})^3/\textbf{k}^2-3(V_s+V_v)\nonumber\\
&&\big
[\psi_{6}(\mathbf{q})(5\omega_1\omega_{1k}+\omega_2\omega_{1k}+\omega_1\omega_{2k}
+\omega_2\omega_{2k}) +\psi_{5}(\mathbf{q})(-5m_1\omega_{1k}+m_2\omega_{1k}+m_1\omega_{2k}-5m_{2}\omega_{2k})\nonumber\\
&&-2\psi_{4}(\mathbf{q})(\omega_1-\omega_2)(\omega_{1k}-\omega_{2k})+\psi_{3}(\mathbf{q})
(-\mathbf{q}^2+m_1m_2-\omega_1\omega_2)(\omega_{1k}+\omega_{2k})\big](\mathbf{k}\cdot
\mathbf{q})^2
  \nonumber\\&&2(V_s-V_v)\big[\psi_{6}(\mathbf{q})(5m_2\omega_1+5m_1\omega_2)-2\psi_{4}(\mathbf{q})(m_2\omega_1+m_1\omega_2)
   +\psi_{5}(\mathbf{q})(-\mathbf{q}^2-5m_1m_2-5\omega_1\omega_2) \nonumber\\
&&+2\psi_{3}(\mathbf{q})(2\mathbf{q}^2+m_1m_2+\omega_1\omega_2)\big](m_2\omega_{1k}+m_1\omega_{2k})\mathbf{k}\cdot \mathbf{q}-\nonumber\\
&&\mathbf{k}^2\mathbf{q}^2(V_s+V_v)\big [
\psi_{6}(\mathbf{q})(5\omega_1\omega_{1k}+\omega_2\omega_{1k}+\omega_1\omega_{2k}
+\omega_2\omega_{2k}) +\nonumber\\
&&\psi_{5}(\mathbf{q})(-5m_1\omega_{1k}+m_2\omega_{1k}+m_1\omega_{2k}-5m_{2}\omega_{2k})
-2\psi_{4}(\mathbf{q})(\omega_1-\omega_2)(\omega_{1k}-\omega_{2k})+\nonumber\\
&&\psi_{3}(\mathbf{q})(-\mathbf{q}^2+m_1m_2-\omega_1\omega_2)(\omega_{1k}+\omega_{2k})\big]\big\}
/[6\omega_1\omega_2(m_2\omega_{1k}+m_1\omega_{2k})].
\end{eqnarray}

\section{$0^+$}
\textbf{normalization condition}
\begin{eqnarray}
\int\frac{d\mathbf{q}}{2\pi^3}\frac{16\psi_2\psi_4
\omega_1\omega_2(-\textbf{q}^2+m_1m_2+\omega_1\omega_2)}{(m_1+m_2)(\omega_1
m_2+\omega_2 m_1)|\textbf{q}|}=2P_{i0}
\end{eqnarray}
 \textbf{Coupled equations }
\begin{eqnarray}
&&\frac{2[\psi_{2}(\mathbf{q})(m_2\omega_{1}+m_1\omega_{2})-\psi_{4}(\mathbf{q})(m_2+m_1)|\textbf{q}|]}{m_2\omega_{1}+m_1\omega_{2}}
=\int\frac{d\mathbf{k}}{(2\pi)^3}
 \{(V_s+V_v)\nonumber\\&&[\psi_{2}(\mathbf{k})(-\mathbf{q}^2+m_2m_1-\omega_1\omega_2)(m_2\omega_{1k}+m_1\omega_{2k})
-\psi_{4}(\mathbf{k})\mathbf{k}(m_2{\omega}_c-m_1{\omega}_s)(\omega_{1k}-\omega_{2k})]\mathbf{k}^2\nonumber\\&&
+(V_s-V_v)[\psi_{2}(\mathbf{k})((m_1-m_2)(\omega_{1k}-\omega_{2k})\mathbf{k}^2+2m_1m_2(m_2\omega_{1k}+m_1\omega_{2k}))-
\nonumber\\&&\psi_{4}(\mathbf{k})|\mathbf{k}|(\omega_1+\omega_2)(m_2\omega_{1k}+m_1\omega_{2k})]\mathbf{k}\cdot
\mathbf{q} \}
/[\mathbf{k}^2\omega_1\omega_2(m_2\omega_{1k}+m_1\omega_{2k})],
\end{eqnarray}
\begin{eqnarray}
&&\frac{2[\psi_{2}(\mathbf{q})(m_2\omega_{1}+m_1\omega_{2})+\psi_{4}(\mathbf{q})(m_2+m_1)|\textbf{q}|]}{m_2\omega_{1}+m_1\omega_{2}}
=\int\frac{d\mathbf{k}}{(2\pi)^3}
\{(V_s+V_v)\nonumber\\&&[\psi_{2}(\mathbf{k})(-\mathbf{q}^2+m_2m_1-\omega_1\omega_2)(m_2\omega_{1k}+m_1\omega_{2k})
+\psi_{4}(\mathbf{k})|\mathbf{k}|(m_2{\omega}_c-m_1{\omega}_s)(\omega_{1k}-\omega_{2k})]\mathbf{k}^2\nonumber\\&&
+(V_s-V_v)[\psi_{2}(\mathbf{k})((m_1-m_2)(\omega_{1k}-\omega_{2k})\mathbf{k}^2+2m_1m_2(m_2\omega_{1k}+m_1\omega_{2k}))+\nonumber\\&&
\psi_{4}(\mathbf{k})|\mathbf{k}|(\omega_1+\omega_2)(m_2\omega_{1k}+m_1\omega_{2k})]\mathbf{k}\cdot
\mathbf{q} \}
/[\mathbf{k}^2\omega_1\omega_2(m_2\omega_{1k}+m_1\omega_{2k})].
\end{eqnarray}

where ${\omega }_{1k}=\sqrt{m^2_{1}+\mathbf{k}^2}$ and ${\omega
}_{2k}=\sqrt{m^2_{q}+\mathbf{k}^2}$; $V_v$ and $V_s$ were given by
ref.\cite{bs}.

\end{document}